\documentstyle[aps,prb]{revtex}

\begin{document}
\draft

\title{Hysteretic current-voltage characteristics and resistance 
switching at a rectifying Ti/Pr$_{0.7}$Ca$_{0.3}$MnO$_{3}$ interface}

\author{A. Sawa,\cite{a,b} T. Fujii,\cite{b,c} M. Kawasaki,\cite{c} 
and Y. Tokura\cite{d}}
\address{Correlated Electron Research Center (CERC),
National Institute of Advanced Industrial Science and Technology (AIST),
Tsukuba, Ibaraki 305-8562, Japan}

\date{\today}
\maketitle

\begin{abstract}
    
We have characterized the vertical transport properties 
of epitaxial layered structures composed of Pr$_{0.7}$Ca$_{0.3}$MnO$_{3}$ 
(PCMO) sandwiched between SrRuO$_{3}$ (SRO) bottom electrode 
and several kinds of top electrodes such as SRO, Pt, Au, Ag, and Ti.
Among the layered structures, Ti/PCMO/SRO is distinct due to 
a rectifying current-voltage ($I$--$V$) characteristic 
with a large hysteresis.
Corresponding to the hysteresis of the $I$--$V$ characteristics, 
the contact resistance of the Ti/PCMO interface reversibly switches
between two stable states by applying pulsed voltage stress.
We propose a model for the resistance switching at the Ti/PCMO 
interface, in which the width and/or height of a Schottky-like barrier 
are altered by trapped charge carriers in the interface states.

\end{abstract}

\pacs{75.30.Vn}



Perovskite manganites have attracted 
considerable interest due to unusual electronic and magnetic 
properties, such as colossal magnetoresistance (CMR),\cite{helmolt,tokura} 
half metallicity,\cite{park} 
and electric field induced switching of resistance, i.e. colossal 
electroresistance (CER).\cite{asamitsu}
Among these characteristics, CER has recently been 
explored in detail.\cite{ponnambalam,oshima,liu,baikalov}
Liu $et$ $al.$, reported that reversible resistance change can be 
induced by applying pulsed voltage at room temperature for 
Pr$_{0.7}$Ca$_{0.3}$MnO$_{3}$(PCMO) sandwiched between Ag top 
electrode and YBa$_{2}$Cu$_{3}$O$_{7}$ or Pt bottom electrode.\cite{liu}
Similar resistance switching phenomena have also been reported in a wide 
variety of perovskite oxides, such as titanates and zirconates.
\cite{blom,beck,watanabe,schmehl,contreras}
This resistance switching has a potential for device applications, such 
as nonvolatile resistance random access memories (RRAM).
Although the resistance switching has been attributed in a previous 
report to the bulk of PCMO,\cite{liu} the resistance value of PCMO 
sandwich structures is much larger than that estimated from the device 
configuration and the resistivity of PCMO bulk crystal.
Baikalov $et$ $al.$\cite{baikalov} 
recently pointed out from multi-leads resistance measurements 
that the resistance switching takes place at the interface between Ag 
electrode and PCMO.
They also proposed the driving mechanism of the switching to be 
electrochemical migration, although experimental evidence was not shown.
Thus, the mechanism of the resistance switching is still unidentified.

Here, we report on the interface properties between PCMO and 
several kinds of metallic electrode such as SrRuO$_{3}$(SRO), 
Pt, Au, Ag, and Ti, in detail, 
through vertical transport measurements of metal/PCMO/SRO
layered structures grown on (100) SrTiO$_{3}$ (STO) single-crystal substrates.
The Ti/PCMO/SRO layered structure, which has 
the largest interface resistance among the tested layered structures, 
is the only one that shows the resistance switching.
The switching behavior can be understood by a model for the interface 
considering the Schottky contact with charge-trapping interface states.


The epitaxial PCMO/SRO layered structures were deposited 
on buffered-HF treated STO substrates\cite{kawasaki} by a pulsed laser 
deposition technique.
Because the SRO film can be epitaxially grown on the STO substrate 
in a step-flow growth mode, the surface of the SRO bottom layer is 
atomically flat.
This allows us to obtain structurally well-defined PCMO/SRO interfaces.
During the deposition, a substrate temperature was kept at 700 
$^{\circ}$C under an oxygen pressure of 100 mTorr.
The thickness of PCMO and SRO layers was 100 nm and 80 nm, 
respectively.
For the sample with an epitaxial SRO layer as the top electrode ($TE$), 
20 nm-thick SRO layer was subsequently deposited on the PCMO layer.
After the depositions, the layered structures were annealed at 400 
$^{\circ}$C for 30 minutes under an oxygen pressure of 500 Torr,
and then cooled down to room temperature.
The crystal structure was analyzed by a four-circle x-ray 
diffractometer.
The full-width at half of maximum of the rocking curves for all 
epitaxial layers in a heterostructure measured together is as narrow as 
0.06 $^{\circ}$.

The device structure is depicted in the inset of Fig. \ref{fig-1}.
Normal metal $TE$ layers were ex-situ deposited by electron beam or 
thermal evaporation under a base pressure of $< 4 \times 10^{-7}$ Torr.
The thickness of the Au and Ag layers is 400 nm, and that of 
the Pt and Ti layers is 20 nm.
For the $TE$s of SRO, Pt and Ti, a 380 nm-thick Au cover layer 
was subsequently deposited.
The layered structures were patterned into mesa structures by 
conventional photolithography and Ar ion milling.


The current--voltage ($I$--$V$) characteristics were measured 
by a three-point contact method.
The size of current and voltage contact pads for the SRO bottom 
electrode (left and right ends in the inset of Fig. \ref{fig-1}, respectively) 
is 0.5 $\times$ 4 mm$^2$ and that of tested junctions (middle) is 200 
$\times$ 200 $\mu$m$^2$.
The use of the separate current and voltage contacts for the SRO 
bottom electrode allows us to extract the 
voltage drop between the SRO bottom electrode and the $TE$ of the 
tested junction.
The reason for using $TE$/PCMO stack for voltage contact pads 
is merely to simplify the device fabrication process.
The positive bias is defined by the current flowing from 
the SRO bottom electrode to the $TE$.
The voltage bias was scanned as 
${0 \to +V_{\rm max} \to 0 \to -V_{\rm max} \to 0}$ V.
All measurements were performed at room temperature.
Figure \ref{fig-1} shows typical $I$--$V$ characteristics of the layered 
structures for five different $TE$s.
The $I$--$V$ characteristic of the epitaxial SRO $TE$ sample shows 
an ohmic behavior, and the resistance ($R_{\rm SRO} \approx 6.8 \Omega$) 
is the smallest among the samples.
For the normal metal $TE$ samples, the resistance ($R_{TE}$) of the 
samples increases in the order of $TE$ = Pt, Au, Ag, and Ti.
From the device configuration and the resistivity of PCMO 
and $TE$ metals, the resistance of PCMO and $TE$ layers along 
vertical direction is estimated 
to be less than 0.1 $\Omega$ and 10 $\mu\Omega$, respectively.
Thus, most of the measured resistance comes from a sum of contact 
resistances at the $TE$/PCMO and PCMO/SRO interfaces and the 
spread resistance of 80 nm-thick SRO bottom electrodes (roughly 10 
$\Omega$).
Therefore, the resistance for the SRO/PCMO/SRO junctions is dominated 
by the spread resistance and the contact resistance at PCMO/SRO 
interface is negligibly small.
In order to estimate the contact resistance for $TE$/PCMO interfaces, 
the difference of the resistance $\Delta R_{TE}$ ($= R_{TE} - R_{\rm SRO}$) 
between the normal metal $TE$ and the SRO $TE$ samples is evaluated, 
and is shown in the inset of Fig. \ref{fig-1}.
The $\Delta R_{TE}$ provides a lower bound for the contact 
resistances for the  normal metal/PCMO interfaces.
For the Ti $TE$ layered structure, we define $R_{\rm Ti}$ as the 
dynamic resistance at $V$ = 0 V because of the nonlinear $I$--$V$ 
characteristic.
The results clearly show that the contact resistances of 
$TE$/PCMO ($TE$ = Pt, Au, Ag, and Ti) interfaces are much larger than the 
resistance of PCMO and $TE$ metal layers themselves, and transport 
properties are dominated by that at the $TE$/PCMO interfaces.

In addition to the nonlinearity, the $I$--$V$ characteristic of the Ti/PCMO 
interface exhibits at the initial state a hysteresis in a negative bias region
but not in a positive bias region.
When the voltage (${|V_{\rm max}| = 2}$ V) was repeatedly biased, 
the current gradually decreased and a hysteresis started to be observed 
both in positive and negative bias regions, as shown in Fig. 
\ref{fig-2}.
The decrease of the current is remarkable in a negative bias region, 
as shown in the inset, 
and finally the Ti/PCMO interface shows a rectifying characteristic.
When much larger voltage (${|V_{\rm max}| = 5}$ V) was applied 
to the virgin Ti/PCMO interface, the current suddenly decreases at a certain 
voltage in a negative bias region, i.e., the $I$--$V$ characteristic has 
a negative dynamic resistance, as denoted by the triangle (${\triangle}$) 
in Fig. \ref{fig-3} (a).
After applying a large negative voltage bias, the Ti/PCMO interface 
shows a rectifying characteristic and a large hysteresis is observed both 
at positive and negative bias, as shown in Fig. \ref{fig-3} (b).
We call hereafter such a bias stress process to get the stable state with 
rectifying and hysteretic characteristics as the $forming$ $process$.

After the forming process, the resistance switching takes place upon the 
application of pulsed voltage stress ($V_{\rm p}$), 
as shown in Fig. \ref{fig-4} (b).
The resistance switching characteristic depends on the initial resistance 
state ($R_{\rm h}^0$ or $R_{\rm l}^0$)
and the pulsed voltage duration (${\tau _{\rm p}}$).
We first define the respective high and low resistance values at a 
voltage bias ($V_{\rm bias}$) of 0.1 V, which were 
obtained after the $I$--$V$ measurements of ${0 \to +5 \to 0 
\to -5 \to 0}$ V and ${0 \to +5 \to 0}$ V, as $R_{\rm h}^0$ and $R_{\rm 
l}^0$ (Fig. \ref{fig-4} (a)).
These correspond to the limiting values for 
$\tau_{\rm p} \to \infty$.
Figure \ref{fig-4} (c) shows $\tau_{\rm p}$ 
dependence of the values ($R_{\rm h}$ and $R_{\rm l}$) for the high 
and low resistance states after the application of 
$V_{\rm p} =$ -5 and + 5 V, respectively.
When the resistance switching is started from $R_{\rm h}^0$, 
the resistance is switched between the $R_{\rm h} \simeq R_{\rm h}^0$ and 
the $R_{\rm l}$, the latter of which decreases with increasing $\tau_{\rm p}$.
When the resistance switching is started from 
the $R_{\rm l}^0$, both of the $R_{\rm h}$ and the $R_{\rm l}$ increase 
with increasing ${\tau _{\rm P}}$ and finally overlap with those started 
from the $R_{\rm h}^0$.
As shown in Fig. \ref{fig-4} (d), 
the ratio of the resistance change ($R_{\rm h}/R_{\rm l}$) increases 
with increasing ${\tau _{\rm P}}$, but the value of $R_{\rm h}/R_{\rm 
l}$ even at $\tau_{\rm p} =$ 0.1 s is less than the one third of 
$R_{\rm h}^0/R_{\rm l}^0$ ($\sim$ 220 k$\Omega$/20 k$\Omega$ = 11).
These results suggest that the high resistance state is much more 
stable than the low resistance one, and that this resistance switching 
is dominated by some slow process.

Here, we first discuss a possible origin of the rectification 
at the Ti/PCMO interface.
The rectifying interface, first of all, reminds us of a Schottky contact.
Assuming that the Ti/PCMO interface is a Schottky contact, 
as shown in the inset of Fig. \ref{fig-3}, from the $I$--$V$ characteristic 
PCMO can be considered as a $p$-type semiconductor .
According to a $p$-type Schottky contact model, a Schottky barrier 
height increases with decreasing the work function of metal electrode.
Because the work function of metals used for $TE$ in this study decreases 
in the order of Pt, Au, and Ag, the $TE$ metal dependence of the 
contact resistance seems to be consistent with a $p$-type Schottky 
contact model. 
However, since Ag and Ti have the almost same work function of $\sim$ 4.3 eV, 
the difference of the interface property between Ag/PCMO and 
Ti/PCMO cannot be understood by a conventional Schottky contact.
Thus, we consider an interface-state-induced band bending at the Ti/PCMO 
interface.
It is well-known that when a density of interface states is high, 
electronic band in a semiconductor bends at an interface, 
independently of a work function of an electrode metal.\cite{sze}
In this model, the degree of band bending, i.e. barrier width 
and height, depends on a net charge in the interface states and an energetical
distribution of those in the band gap, and 
this band bending also causes a rectification.
Because Ti is a getter for oxygen and has a small electronegativity, 
the Ti layer possibly extracts a large amount of oxygen atoms from 
the surface of the PCMO layer.
The high density of the interface states induced by the oxygen vacancies may 
cause a large degree of the band bending at the Ti/PCMO interface, as 
compared with the Ag/PCMO interface.

Next, on the basis of the interface-state-induced band bending picture, 
we propose a possible model of the resistance switching.
By applying large voltage at the Ti/PCMO interface, a large amount of electrons 
is accumulated (extracted) into (from) the interface states upon 
reverse (forward) bias.
Accordingly, a variation of a net charge in the interface states leads to a 
modification of an Schottky-like 
barrier width and/or height, because the degree of the band 
bending depends on a net charge in the interface states.
This model is similar to the one proposed for a resistance 
switching in a layered structure of a ferroelectric oxide.\cite{blom,schmehl}
The sandwich structure of 
Au/PbTiO$_{3}$/La$_{0.5}$Sr$_{0.5}$CoO$_{3}$, where PbTiO$_{3}$ 
was a ferroelectric $n$-type semiconductor, showed a resistance 
switching.\cite{blom}
The mechanism of the resistance switching was explained by the 
change of the Schottky barrier width at the Au/PbTiO$_{3}$ interface caused 
by a polarity alternation of a space charge in the ferroelectric 
PbTiO$_{3}$ layer.

A remaining open question is what happens in the forming process.
One possible candidate is the electrochemical migration of oxygen atoms.
Oxygen gettering by Ti at the Ti/PCMO interface may generate the 
oxygen-defect-induced interface states,
resulting in a Schottky-like barrier.
However, further study is needed for elucidating the origin of the 
forming process.
A way to avoid such chemical problems is to employ heteroepitaxial 
Schottky junctions made of oxide semiconductors and oxide metals.


In summary, we have demonstrated pulsed voltage induced resistance switching 
at the Ti/PCMO interface, which is accompanied by hysteretic and rectifying 
$I$--$V$ characteristics.
The resistance switching can be explained by a model of the interface 
with trapping states which have sufficiently high density to form 
a Schottky-like barrier. 

We would like to thank T. Shimizu, I.H. Inoue, A. Odagawa, H. Yamada, 
J. Matsuno, and H. Akoh for useful discussions.



\begin{figure}
\caption{$I$--$V$ characteristics of $TE$/PCMO/SRO layered structures 
with five different $TE$s. Here, $TE$, PCMO, and SRO stand for top 
electrode, Pr$_{0.7}$Ca$_{0.3}$MnO$_{3}$, and SrRuO$_{3}$, 
respectively. The upper panel of the insets shows 
a schematic of the samples. 
The lower one shows $\Delta R_{TE}$ (= $R_{TE} - R_{\rm SRO}$) for $TE$ = Pt, Au, Ag, and Ti.
\label{fig-1}}
\end{figure}

\begin{figure}
\caption{$I$--$V$ characteristics of a Ti/PCMO/SRO layered 
structure measured with repeated voltage scan (${|V_{\rm 
max}| = 2}$ V) up to 45th cycle. 
The inset shows the time chart of the current.
\label{fig-2}}
\end{figure}

\begin{figure}
\caption{$I$--$V$ characteristics of a Ti/PCMO/SRO layered 
structure drawn in (a) linear and (b) semilogarismic current scales.
Insets schematically show electronic band diagrams for a rectifying Ti/PCMO 
interface. 
\label{fig-3}}
\end{figure}

\begin{figure}
\caption{
(a) Forward bias $I$--$V$ characteristic of a Ti/PCMO/SRO layered structure.
Low and high resistance states ($R_{\rm l}^0$ and $R_{\rm h}^0$) 
after voltage scans of $|V_{\rm max}|$ = 5 V (see text) are defined at 
the filled and open stars, respectively.
Each resistance value was evaluated by measuring current at a $V_{\rm 
bias}$ of 0.1 V.
(b) Resistance switching behavior (bottom) started from the 
$R_{\rm l}^0$ state, by applying a sequence of pulsed voltage stress 
(top) of $V_{\rm p} = \pm 5$ V. 
(c) Pulsed voltage duration ($\tau _{\rm P}$) dependence of the low (square) and high (circle) 
resistance states.
The filled and open symbols represent the data
started from $R_{\rm l}^0$ and $R_{\rm h}^0$, respectively.
(d) $\tau _{\rm P}$ dependence of the resistance ratio 
($R_{\rm h}/R_{\rm l}$).
\label{fig-4}}
\end{figure}

    

\begin{thebibliography}{99}

\bibitem[{\rm a)}]{a} Electronic mail: a.sawa@aist.go.jp

\bibitem[{\rm b)}]{b} Also at: CREST-JST, Japan.

\bibitem[{\rm c)}]{c} Also at: Institute for Materials Research, 
Tohoku University, Sendai 980-8577, Japan.
resistance
\bibitem[{\rm d)}]{d} Also at: Department of Applied Physics, 
University of Tokyo, Tokyo 113-8656, Japan.

\bibitem{helmolt} R. von Helmolt, J. Wecker, B. Holzapfel, L. Schultz, 
and K. Samwer, Phys. Rev. Lett. {\bf 71}, 2331 (1993).

\bibitem{tokura} Y. Tokura and Y. Tomioka, J. Magn. Magn. Mater. 
{\bf 200}, 1 (1999).

\bibitem{park} J.-H. Park, E. Vescovo, H.-J. Kim, C. Kwon, R. Ramesh, 
and T. Venkatesan, Nature (London) {\bf 392}, 794 (1998).

\bibitem{asamitsu} A. Asamitsu, Y. Tomioka, H. Kuwahara, and Y. Tokura,
Nature {\bf 388}, 50 (1997).

\bibitem{ponnambalam} V. Ponnambalam, S. Parashar, A.R. Raju, and C.N.R. 
Rao, Appl. Phys. Lett. {\bf 74}, 206 (1999).

\bibitem{oshima} H. Oshima, K. Miyano, Y. Konishi, M. Kawasaki, and Y. 
Tokura, Appl. Phys. Lett. 
{\bf 75}, 1473 (1999).

\bibitem{liu} S.Q. Liu, N.J. Wu, and A. Ignatiev, Appl. Phys. Lett. 
{\bf 76}, 2749 (2000).

\bibitem{baikalov} A. Baikalov, Y.Q. Wang, B. Shen, B. Lorenz, S. 
Tsui, Y.Y. Sun, Y.Y. Xue, and C.W. Chu, Appl. Phys. Lett. {\bf 83}, 
957 (2003). 

\bibitem{blom} P.W.M. Blom, R. M. Wolf, J.F.M. Cillessen, and M.P.C.M. 
Krijn, Phys. Rev. Lett. {\bf 73}, 2107 (1994).

\bibitem{beck} A. Beck, J.G. Bednorz, Ch. Gerber, C. Rossel, and D. 
Widmer, Appl. Phys. Lett. {\bf 77}, 139 (2000)

\bibitem{watanabe} Y. Watanabe, J.G. Bednorz, A. Bietsch, Ch. Gerber, 
D. Widmer, A. Beck, and S.J. Wind, Appl. Phys. Lett. {\bf 78}, 3738 
(2001).

\bibitem{schmehl} A. Schmehl, F. Lichtenberg, H. Bielefeldt, J. 
Mannhart, and D.G. Schlom, Appl. Phys. Lett. {\bf 82}, 3077 (2003).

\bibitem{contreras} J.R. Contreras, H. Kohlstedt, U. Poppe, R. Waser, C. 
Buchal, and N.A. Pertsev, Appl. Phys. Lett. {\bf 83}, 4595 (2003).

\bibitem{kawasaki} M. Kawasaki, K. Takahashi, T. Maeda, R. Tsuchiya, 
M. Shinohara, O. Ishiyama, T. Yonezawa, M. Yoshimoto, and H. Koinuma, 
Science {\bf 266}, 1504 (1994).

\bibitem{sze} S.M. Sze, {\it Physics of Semiconductor Devices, 2nd 
ed.} (Wiley, New York, 1981).

\end{thebibliography}
\end{document}